\documentclass[usenatbib,useAMS]{mn2e}
\usepackage[fleqn,tbtags]{amsmath} 
\usepackage{times,aas_macros,amssymb}
\usepackage{natbib,graphicx}

\pubyear{2010} 
\title[Reionization in overdense regions]{Reionization
  and feedback in overdense regions at high redshift} 
\author[Kulkarni \& Choudhury]
{Girish Kulkarni\thanks{E-mail: girish@hri.res.in}~
and
T. Roy Choudhury\thanks{E-mail: tirth@hri.res.in}\\
Harish-Chandra Research Institute, Chhatnag Road, Jhunsi, Allahabad 211019, India
} 
\date{}

\begin{document}
\maketitle

\begin{abstract}
Observations of galaxy luminosity function at high redshifts typically
focus on fields of view of limited sizes preferentially containing
bright sources. These regions possibly are overdense and hence biased
with respect to the globally averaged regions.  Using a semi-analytic
model based on Choudhury \& Ferrara (2006) which is calibrated to
match a wide range of observations, we study the reionization and
thermal history of the universe in overdense regions. The main results
of our calculation are: (i) Reionization and thermal histories in the
biased regions are markedly different from the average ones because of
enhanced number of sources and higher radiative feedback. (ii) The
galaxy luminosity function for biased regions is markedly different
from those corresponding to average ones.  In particular, the effect
of radiative feedback arising from cosmic reionization is visible at
much brighter luminosities.  (iii) Because of the enhanced radiative
feedback within overdense locations, the luminosity function in such
regions is more sensitive to reionization history than in average
regions.  The effect of feedback is visible for absolute AB magnitude
$M_{AB} \gtrsim -17$ at $z=8$, almost within the reach of present day
observations and surely to be probed by the James Webb Space Telescope
(JWST).  This could possibly serve as an additional probe of radiative
feedback and hence reionization at high redshifts.
\end{abstract}

\section{Introduction}
Deep surveys have now discovered galaxies at redshifts close to the
end of reionization \citep{2006NewAR..50..152B, 2006Natur.443..186I,
  2007ApJ...670..928B, 2007ApJ...656L...1H, 2007ApJ...663...10S,
  2008ApJ...686..230B, 2008ApJ...678..647B, 2008ApJ...680L..97H,
  2008ApJ...677...12O, 2008ApJ...685..705R, 2009arXiv0909.2255B,
  2009ApJ...705..936B, 2009ApJ...697.1128H, 2009MNRAS.395.2196M,
  2009ApJ...690.1350O, 2009ApJ...696.1164O, 2009arXiv0912.4263B,
  2009ApJ...706.1136O, 2009MNRAS.398L..68S, 2009ApJ...697.1907Z,
  2010ApJ...709L..16O, 2010A&A...511A..20C, 2010arXiv1003.1706B,
  2010MNRAS.404..212H, 2010MNRAS.403..960M}.  Luminosity function of
these galaxies, and its evolution, can answer important questions
about reionization.  Indeed, much work has been done on constructing
self-consistent models of structure formation and the evolution of
ionization and thermal state of the IGM that explain these
observations \citep{2005MNRAS.361..577C, 2005ApJ...623..627H,
  2005ApJ...625....1W, 2006MNRAS.371L..55C, 2007MNRAS.379..253D,
  2007MNRAS.377..285S, 2008MNRAS.391...63I, 2009MNRAS.398.2061S}.
Studies of the Gunn-Peterson trough \citep{1965ApJ...142.1633G} at
$z\geq 6$ have established that the mean neutral hydrogen fraction is
higher than $10^{-4}$ (e.~g.\ \citealt{2006AJ....132..117F}) and it is
most likely that the IGM is still highly ionized at these redshifts
\citep{2008MNRAS.386..359G, 2008MNRAS.388L..84G}.  Furthermore, CMB
observations indicate the electron scattering optical depth to the
last scattering surface to be $\tau_e=0.088\pm 0.015$ based on the
WMAP seven year data.  A combination of high redshift luminosity
function data with the data from these absorption systems and CMB
observations favour an extended epoch of reionization that begins at
$z\approx 20$ and ends at $z\approx 6$ \citep{2006MNRAS.371L..55C}.

Nonetheless, interpreting high redshift luminosity functions is not
straightforward and detailed modelling is required.  For instance,
local HII regions around these galaxies can affect luminosity function
evolution \citep{2005ApJ...621...89C} and clustering of galaxies can
enhance this effect \citep{2005astro.ph..7014C}.  Another complication
is because of the fact that these surveys can detect only the
brightest galaxies at these high redshifts ($z\gtrsim 6$).  Such
galaxies can form only in highly overdense regions and therefore the
surveyed volume is far from average.  An important question in that
case is whether reionization proceeds differently in such regions
\citep{2007MNRAS.375.1034W}.  

Galaxy formation is enhanced in overdense regions because of a
positive bias in abundance of dark matter haloes.  The enhancement in
the number of galaxies is proportional to the mass overdensity in the
region, with the constant of proportionality (`bias') related to halo
masses and collapse redshifts \citep{2002PhR...372....1C}.  This
increases the number density of sources of ionising radiation and aids
reionization of the intergalactic medium (IGM) in overdense regions.
However, an increase in the IGM density also adds to radiative
recombination.  Furthermore, reionization is accompanied by radiative
feedback \citep{1996ApJ...465..608T}.  Radiative feedback heats the
IGM and suppresses formation of low mass galaxies.  This increase in
radiative recombinations and feedback works against the process of
reionization and the two effects need not cancel out.  Relative
significance of these negative and positive contributions will
determine how differently reionization evolves in overdense regions.

\begin{figure*}
\begin{center}
\begin{tabular}{cc}
\includegraphics[scale=0.5]{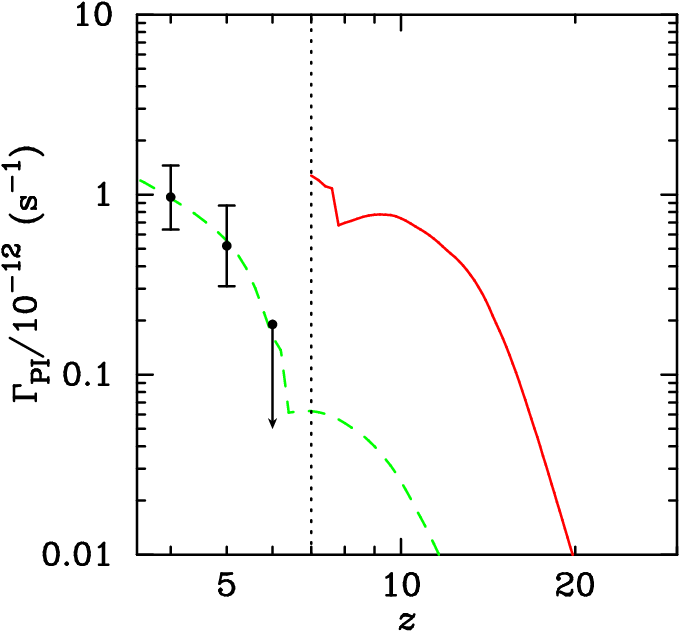} &
\includegraphics[scale=0.5]{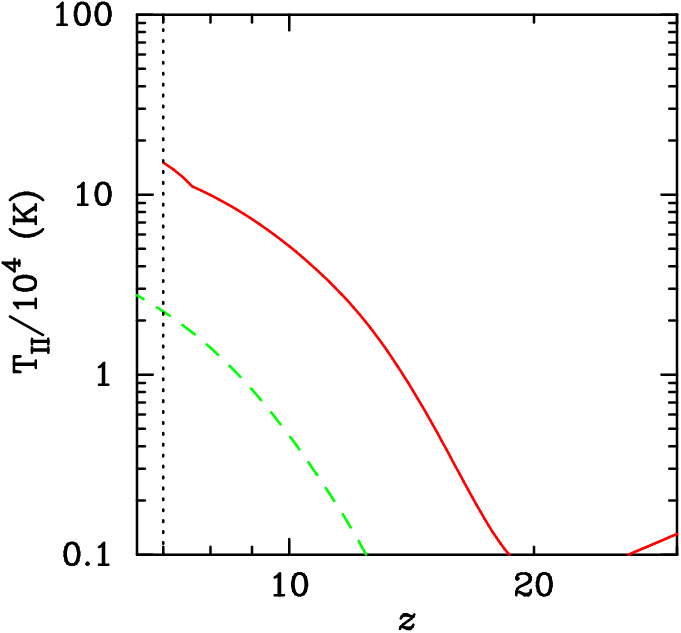} \\
\includegraphics[scale=0.5]{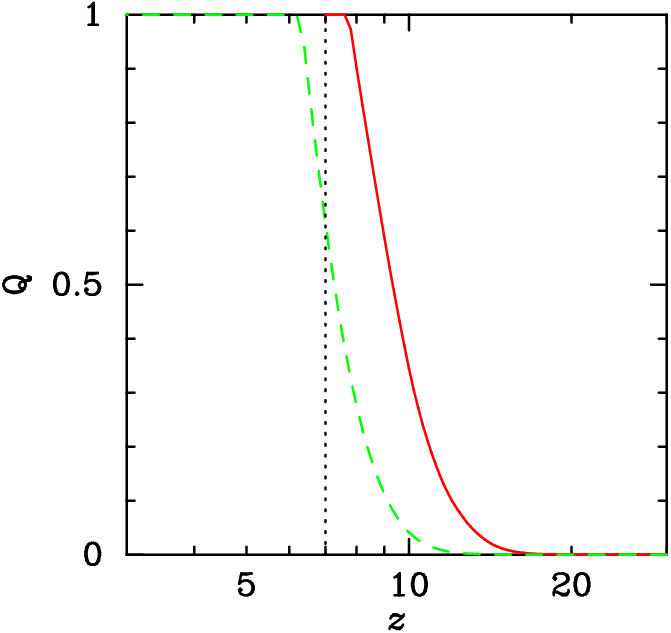} &
\includegraphics[scale=0.5]{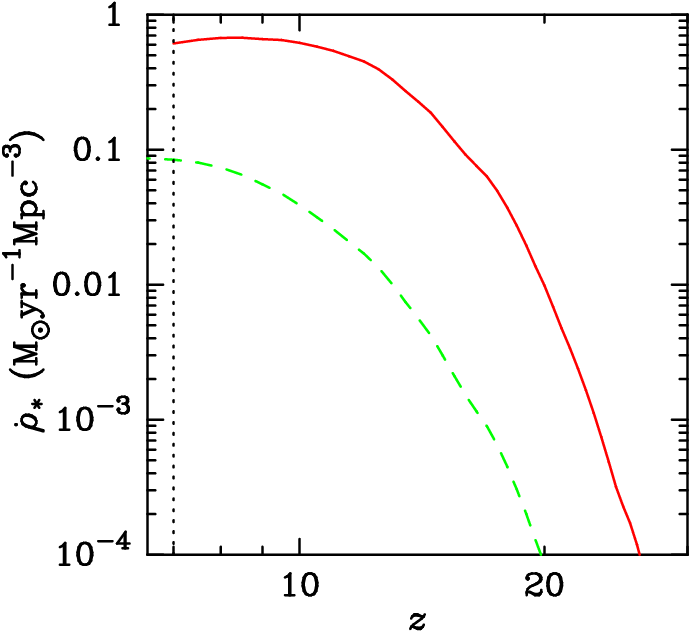} \\
\end{tabular}
\end{center}
\caption{Behaviour of various quantities in our fiducial model, in the
  average and overdense regions are shown by dashed and solid lines
  respectively. The top left panel shows the photoionisation rate,
  with data points taken from \citep{2007MNRAS.382..325B}. The top
  right panel shows the mass-averaged temperature for ionised regions,
  which essentially determine the radiative feedback. The bottom left
  panel is for the volume filling factor of ionised regions. The
  bottom right panel shows the cosmic star formation rate.  Note that
  the overdense region that we consider here collapses at $z=6.8$.
  The vertical dotted line in the top left panel highlights this.  We
  cannot evolve our reionization model for smaller redshifts.}

\label{BestFitReionizationModel}
\end{figure*}

Recently, \citet{2009ApJ...695..809K} studied a sample of
$i_{775}$-dropout candidates identified in five Hubble Advanced Camera
for Surveys (ACS) fields centred on Sloan Digital Sky Survey (SDSS)
QSOs at redshifts $z\approx 6$.  They compared results with those from
equally deep Great Observatory Origins Deep Survey (GOODS)
observations of the same fields in order to find an enhancement or
suppression in source counts in ACS fields.  An enhancement would
imply that bias wins over negative feedback in these overdense
regions.  They found the ACS populations to be overdense in two
fields, underdense in two field, and equally dense as the GOODS
populations in one field.  Somewhat surprisingly, they did not find a
clear correlation between density of $i_{775}$ dropouts and the
region's overdensity.  

In this paper, we use semi-analytic models to study reionization
within overdense regions. The main aim of this work is to quantify the
effects of enhancement in the number of sources and radiative feedback
within such regions and explore the possibility whether the galaxy
luminosity function in overdense regions can be used as potential
probe of feedback and reionization history.  It is known that if
ionization feedback is the main contributor to the suppression of star
formation in low mass haloes then one can distinguish between early
and late reionization histories by constraining the epoch at which
feedback-related low-luminosity flattening occurs in the galactic
luminosity function.  The effect of reionization feedback on the high
redshift galaxy luminosity function was first demonstrated using
semi-analytic models by \citet{2007MNRAS.377..285S}.  We apply their
method to study the luminosity function in overdense regions.

We describe our model for reionization and give details about various
parameters and their calibration in \S 2.  We explain our method of
identifying biased regions and outline modification to our
reionization model in such regions in \S 3.  Details of our luminosity
function calculation appear in \S 4.  We present and discuss our
results in \S 5 and \S 6.  Throughout the paper, we use the best-fit
cosmological parameters from the 7-year WMAP data
\citep{2010arXiv1001.4635L}, i.e., a flat universe with $\Omega_m =
0.26$, $\Omega_{\Lambda} = 0.73$, $\Omega_K = 0.044$ and $\Omega_b = 0.045$, and $h=0.713$.  The parameters defining the linear dark matter power spectrum are $\sigma_8=0.80$, $ n_s=0.96$, ${\rm d} n_s/{\rm d} \ln k=-0.034$.

\section{Description of the Semi-analytic model}

In this section, we first summarise the basic features of the
semi-analytic model used for studying the globally averaged
reionization history. We then describe in detail the modifications
made to this model in order to study reionization in biased regions.

\subsection{Globally averaged reionization}

Our model for reionization and thermal history of the average IGM is
essentially that developed in \citealt{2005MNRAS.361..577C} (CF05).
The main features of this model are as follows.

The model accounts for IGM inhomogeneities by adopting a lognormal
distribution with the evolution of volume filling factor of ionized
hydrogen (H\textsc{ii}) regions $Q_{\rm HII}(z)$ being calculated
according to the method outlined in \citet{2000ApJ...530....1M};
reionization is said to be complete once all the low-density regions
(say, with overdensities $\Delta < \Delta_{\rm crit} \sim 60$) are
ionised.  We follow the ionization and thermal histories of neutral
and HII regions simultaneously and self-consistently, treating the IGM
as a multi-phase medium.  In this work, we do not consider the
reionization of singly ionised helium as it occurs much later ($z \sim
3$) than redshifts of our interest.

The number of ionising photons depends on the assumptions made
regarding the sources.  In this work, we have assumed that
reionization of hydrogen is driven by stellar sources.  The rate of
ionising photons injected into the IGM per unit time per unit volume
at redshift $z$ is denoted by $\dot n_\mathrm{ph}(z)$ and is
essentially determined by the star formation rate (SFR) density
$\dot{\rho}_*(z)$.  The first step in this calculation is to evaluate
the comoving number density $N(M,z,z_c)dMdz_c$ at redshift $z$ of
collapsed halos having mass in the range $M$ and $M+dM$ and redshift
of collapse in the range $z_c$ and $z_c+dz_c$
\citep{1994PASJ...46..427S}:
\begin{equation}
\begin{split}
N(M,z,z_c)dMdz_c=N(&M,z_c)\nu^2(M,z_c)\frac{\dot
  D(z_c)}{D(z_c)}\\&\times p_\mathrm{surv}(z,z_c)\frac{dt}{dz_c}dz_cdM,
\end{split}
\label{nmzzc}
\end{equation}
where $N(M,z_c)dM$ is the comoving number density of collapsed halos
with mass between $M$ and $M+dM$, also known as the Press-Schechter
(PS) mass function \citep{1974ApJ...187..425P}, and
$p_\mathrm{surv}(z,z_c)$ is the probability of a halo collapsed at
redshift $z_c$ surviving without merger till redshift $z$.  This
survival probability is simply given by
\begin{equation}
p_\mathrm{surv}(z,z_c)=\frac{D(z_c)}{D(z)},
\end{equation}
where $D(z)$ is growth function of matter perturbations.  Furthermore,
$\nu(M,z_c)$ is given by $\delta_c/[D(z_c)\sigma(M)]$, where
$\sigma(M)$ is the rms value of density fluctuations at the comoving
scale corresponding to mass $M$ and $\delta_c$ is the critical
overdensity for collapse of the halo.  Next, we assume that the SFR of
a halo of mass $M$ that has collapsed at an earlier redshift $z_c$
peaks around a dynamical time-scale of the halo and has the form
\begin{equation}
\begin{split}
\dot M_*(M,z,z_c)=f_*\left(\frac{\Omega_b}{\Omega_m}M\right)&\frac{t(z)-t(z_c)}{t^2_\mathrm{dyn}(z_c)}\\&\times\exp\left[-\frac{t(z)-t(z_c)}{t_\mathrm{dyn}(z_c)}\right].
\end{split}
\label{halo-sfr}
\end{equation}
where $f_*$
denotes the fraction of the total baryonic mass of the halo that gets
converted into stars.  
The global SFR density at redshift $z$ is then
\begin{equation}
\dot\rho_*(z)=\int_z^\infty \!\!dz_c\int_{M_\mathrm{min}(z_c)}^\infty
\!\!dM\dot M_*(M,z,z_c)N(M,z,z_c),
\label{global_sfr}
\end{equation}
where the lower limit of the mass integral, $M_\mathrm{min}(z_c)$,
prohibits low-mass halos from forming stars; its value is decided by
different feedback processes. In this work, we exclusively consider
radiative feedback.  For neutral regions, we assume that this quantity
is determined by atomic cooling of gas within haloes (we neglect
cooling via molecular hydrogen). Within ionised regions, photo-heating
of the gas can result in a further suppression of star formation in
low-mass haloes.  We compute such (radiative) feedback
self-consistently from the evolution of the thermal properties of the
IGM, as discussed in Section \ref{radfb}.

We can then write the rate of emission of ionising photons per unit
time per unit volume per unit frequency range, $\dot n_\nu(z)$, as
\begin{equation}
\dot n_\nu(z)=N_\gamma(\nu) f_\mathrm{esc}\dot\rho_*(z),
\label{nnu}
\end{equation}
where $N_\gamma(\nu)$ is the total number of ionising photons emitted
per unit frequency range per unit stellar mass and $f_\mathrm{esc}$ is
the escape fraction of photons from the halo.  The quantity
$N_\gamma(\nu)$ can be calculated using population synthesis, given
the initial mass function and spectra of stars of different masses
\citep{2007MNRAS.377..285S}.  In this paper we have used the
population synthesis code Starburst99 \citep{1999ApJS..123....3L,
  2005ApJ...621..695V} to calculate $N_\gamma(\nu)$ by evolving a
stellar population of total mass $10^6$ M$_\odot$ with a $0.1-100.0$
M$_\odot$ Salpeter IMF and metallicity $0.001$ ($0.05$ times the solar
metallicity, $Z_\odot=0.02$).  The total rate of emission of ionising
photons per unit time per unit volume is obtained simply integrating
by Equation (\ref{nnu}) over suitable frequency range.

Given the above model, we obtain best-fit parameters by comparing with
the redshift evolution of photoionisation rate obtained from the
Ly$\alpha$ forest \citep{2007MNRAS.382..325B} and the electron
scattering optical depth \citep{2010arXiv1001.4635L}.  We should
mention here that any model containing only a single population of
atomic-cooled stellar sources with non-evolving $f_* f_{\rm esc}$
cannot match both the Ly$\alpha$ forest and WMAP constraints
\citep{2008MNRAS.385L..58C, 2007MNRAS.382..325B}.  In this work, we
choose the model which satisfies the Ly$\alpha$ constraints but
underpredicts $\tau_e$. In order to match both the constraints, one
has to invoke either molecular cooling in minihaloes and/or metal-free
stars and/or other unknown sources of reionization.  This model is
described by the parameter values $f_*=0.2$ and
$f_\mathrm{esc}=0.135$, and gives $\tau_e$ of 0.072.  Figure
\ref{BestFitReionizationModel} shows evolution of the filling factor
of ionised regions, global star formation rate density, mass-weighted
average temperature in ionised regions and average hydrogen
photoionisation rate in this model (dashed curves in all the panels).
The filling factor of ionized regions is seen to rise monotonically
from $z\approx 15$ and takes values close to unity at redshifts
$z\approx 6$.  Temperature of ionized regions also rises rapidly
during reionization and flattens out to a few times $10^4$ K at
redshift $z\lesssim 4$ (not shown here).  Lastly, the photoionisation
rate also increases during reionization as the star formation rate
builds up.  However, the photoionisation rate increases rapidly with a
sudden jump at $z\approx 6$ when the ionized regions overlap (filling
factor becomes close to unity).  This is because a given region in
space starts receiving ionizing photos from multiple sources and as a
result, the ionizing flux suddenly increases.  This is our fiducial
model, which satisfies observational constraints from Ly$\alpha$
forest, observations of star formation rate history, number density of
Lyman-limit systems at high redshift and of the IGM temperature.  In
this model, reionization starts at $z\approx 15$ and is 90\% complete
by $z\approx 7$.  Evolution of $x_\mathrm{HII}$ is consistent with
constraints from Ly$\alpha$ emitters and the GP optical depths.

\begin{figure*}
\begin{center}
\begin{tabular}{ccc}
\includegraphics[scale=0.5]{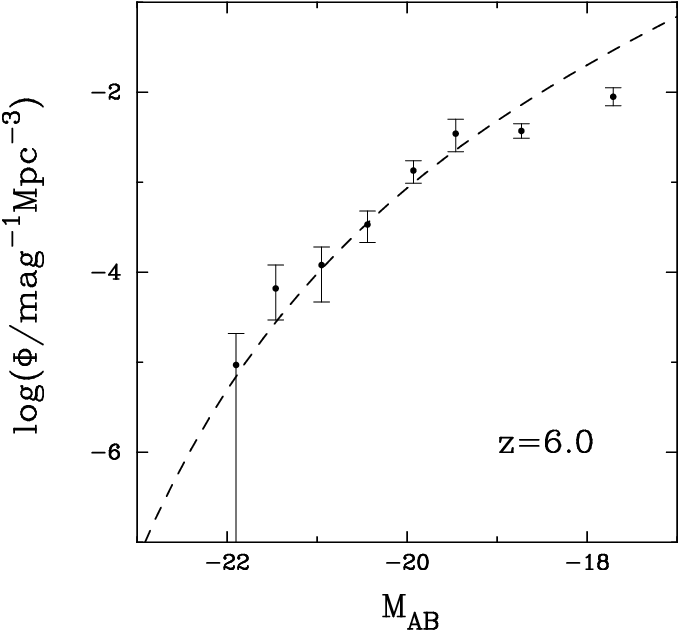} 
\includegraphics[scale=0.5]{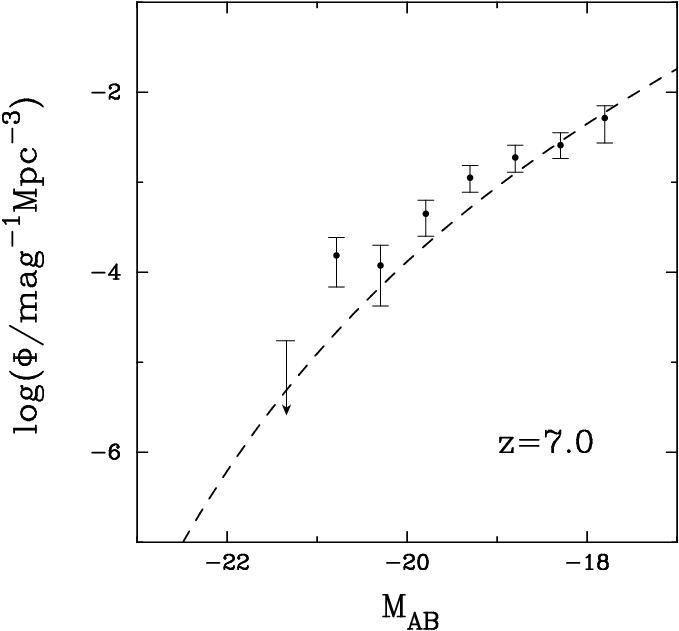} 
\includegraphics[scale=0.5]{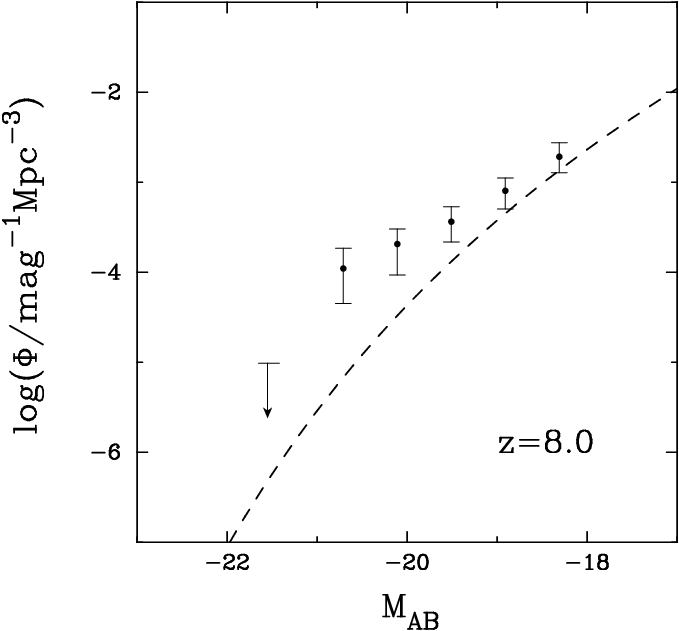} 
\end{tabular}
\end{center}
\caption{Luminosity function from our fiducial model at $z=6$, $7$ and
  $8$.  This is the average case.  Data points are from
  \citet{2006NewAR..50..152B} ($z=6$) and \citet{2010arXiv1006.4360B}
  ($z=7$, $8$).  }
\label{lf}
\end{figure*}

Having set up the reionization model, we then calculate the predicted
luminosity function of galaxies in this model.  Luminosity functions
of objects are usually preferred for comparing theory with
observations because of its directly observable nature.  In this work,
we closely follow the approach presented by
\citet{2007MNRAS.377..285S} to calculate the luminosity function.  We
obtain luminosity per unit mass, $l_{1500}(t)$, at 1500 \AA\ as a
function of time from population synthesis for an instantaneous burst.
In our model, star formation does not happen in a burst, but is a
continuous process spread out over a dynamical time-scale.  Therefore,
in order to determine the luminosity of a halo, $L_{1500}(t)$ with
this kind of star formation, we convolve $l_{1500}(t)$ with the halo's
star formation rate using
\begin{equation}
L_{1500}(M,T)=\int_T^0d\tau\dot
M_\mathrm{*}(M,T-\tau,z_c)l_{1500}(\tau),
\label{HaloL}
\end{equation}
where $T$ is the age of the halo, which has mass $M$ and which
collapsed at redshift $z_c$.  
This luminosity can be converted to
absolute AB magnitude using
\begin{equation}
M_{AB}=-2.5\log_{10}(L_{\nu0})+51.60,
\end{equation}
where the luminosity is in units of erg s$^{-1}$ Hz$^{-1}$
\citep{1983ApJ...266..713O}.  One can compute the luminosity
evolution for any halo that collapses at redshift $z_c$ and undergoes
star formation according to Equation (\ref{halo-sfr}).  The luminosity
function at redshift $z$, $\Phi(M_{AB},z)$, is now given by
\begin{equation}
\begin{split}
\Phi(&M_{AB},z)dM_{AB}\\&=\int_z^\infty dz_c N(M,z,z_c)\frac{dM}{dL_{1500}}\frac{dL_{1500}}{dM_{AB}}dM_{AB},
\end{split}
\label{lumfn}
\end{equation}
where $N(M,z,z_c)$ is the number density at redshift $z$ of halos of
mass $M$ collapsed at redshift $z_c$.  We will use Equation
(\ref{lumfn}) to study effect of overdensity on the luminosity
function and to compare the luminosity function in our model with
observations in the next section.

Figure \ref{lf} shows the globally averaged luminosity function
calculated using our model for different redshifts in comparison with
observations presented by \citet{2006NewAR..50..152B}.  We find that
our model reproduces the observed luminosity functions at high
redshifts reasonably well.  In particular, the match at $z=6$ is
remarkably good while the model predicts less number of galaxies than
what is observed at $z=7$ and 8. This could indicate that the
star-forming efficiency $f_*$ increases with $z$, and/or the
time-scale of star formation is lower than $t_{\rm dyn}$ at higher
redshifts. The match of the model with the data can be improved by
tuning these parameters suitably, however we prefer not to introduce
additional freedom in constraining the parameters; rather our focus is
to estimate the effect of reionization and feedback on the luminosity
function.

In our calculation of luminosities, we do not make any correction for
dust.  This is partly because of indications from observed very blue
UV-continuum slopes \citep{2010ApJ...708L..69B,2010ApJ...709L..16O,
  2010ApJ...719.1250F,2010MNRAS.tmp.1378B} that dust extinction in
$z\gtrsim 7$ is small.  As discussed in the next section, we
exclusively work with luminosity functions at these redshifts.  Also,
the effect of dust is degenerate with $f_*$ to some extent.
Therefore, the exclusion of dust extinction does not affect the
general results of our calculation.  

\begin{figure}
\begin{center}
\includegraphics[scale=0.65]{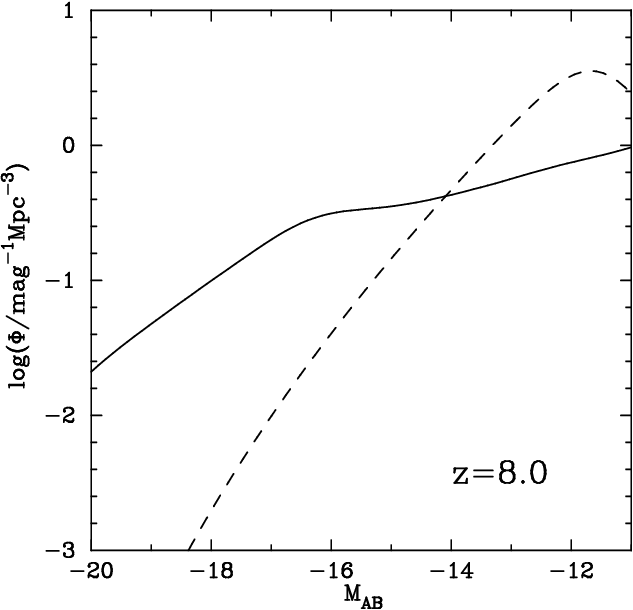} 
\end{center}
\caption{Effect of overdensity on the luminosity function via feedback
  at z=8.0.  This is for $\delta=8.8$ and $R_L=1.482$ Mpc.  }
\label{lf78}
\end{figure}

\subsection{Biased regions}
\label{biassec}

We have mentioned that galaxy luminosity functions provide valuable
information regarding reionization.
However, observations are carried out over relatively small fields of
view.  
The bright sources in these fields are typically hosted by high mass
haloes.
Hence, it is most likely that these fields are biased tracers of
luminosity function and reionization.
In this section, we extend our model to study reionization within such
biased regions and quantify the departure of various quantities from
their globally averaged trends.
Reionization in biased regions has been discussed in the literature.
\citet{2007MNRAS.375.1034W} studied the correlation between high
redshift galaxy distribution and the neutral Hydrogen 21 cm emission
by considering reionization in the vicinity of these galaxies.
\citet{2008MNRAS.383..691W} considered the ionization background near
high redshift quasars.  
\citet{2009MNRAS.399.1877G} studied effect of reionization around high
redshift quasars on the power spectrum of 21 cm emission (see also
\citealt{2007MNRAS.376.1680P}). 
The general conclusion of these studies is that overdense regions are
ionised earlier.  
In this work, we will consider the behaviour of luminosity functions
in such regions.

Overdense regions are characterised by their comoving Lagrangian size
$R$, and their linearly extrapolated overdensity $\delta$.
At a given redshift, we can take a scale $R$ corresponding to an
observed field of view (e.g., WFC3/IR field in HST) and then determine
$\delta$ by identifying the presence of a massive object, for example,
a quasar or a bright galaxy.
We follow a prescription discussed by \citet{2008MNRAS.385.2175M}.

Note that if a galaxy with luminosity $L_{1500}$ is observed at
redshift $z_g$ then we can assign a certain mass to the dark matter
halo containing the galaxy, say $M$.
The halo mass $M$ has to be obtained from galaxy luminosity $L_{1500}$
by using some prescription or by fitting the galaxy's spectral energy
distribution (SED; \citealt{2004MNRAS.353..189V}).
In this work, however, since uncertainty in the value of
overdensity $\delta$ is expected to be larger than any uncertainty in
the mass of the galaxy's host halo, we choose to calculate this mass
in an alternate, simpler manner.
Note that we can invert equations (\ref{HaloL}) and (\ref{halo-sfr})
to obtain $M$ given an observed value of $L_{1500}$ if we have an
estimate of the redshift at which the galaxy formed.
In fact, we obtain the halo mass $M$ such that the halo luminosity
after a dynamical time from halo formation time, calculated according
to our model, is equal to $L_{1500}$.
In other words, we break the degeneracy between halo mass and
formation redshift by assuming that the galaxy's age is equal to the
dynamical time of the halo.

From this analysis, we conclude that a collapsed object with mass $M$
exists at redshift $z_g$.  Now suppose our field of observation
corresponds to some linear scale $R_o$ at this redshift.  Then the
linearly extrapolated overdensity at this scale can be obtained using
the excursion set prescription \citep{2008MNRAS.385.2175M}.  Recall
that the probability distribution of the extrapolated Gaussian density
field smoothed over scale $R$ is also Gaussian
\begin{equation}
  Q(\delta_R,\sigma^2(R))d\delta_R=\frac{1}{\sqrt{2\pi \sigma^2(R)}}\exp\left[-\frac{\delta_R^2}{2\sigma^2(R)}\right]d\delta_R.
\end{equation}
The conditional probability distribution of overdensity $\delta_1$ on
a scale specified by the variance $\sigma^2_1$, given a value of
overdensity $\delta_2$ on a larger scale specified by the variance
$\sigma^2_2<\sigma^2_1$ is given by
\begin{equation}
  Q(\delta_1,\sigma^2_1|\delta_2,\sigma^2_2)=Q(\delta_1-\delta_2,\sigma^2_1-\sigma^2_2).
\end{equation}
Conversely, when the value of overdensity $\delta_1$ on a smaller
scale specified by variance $\sigma^2_1$ is given, the conditional
probability distribution of $\delta_2$ can be obtained using Bayes
theorem as 
\begin{equation}
  Q(\delta_2,\sigma^2_2|\delta_1,\sigma^2_1)\propto
  Q(\delta_1,\sigma^2_1|\delta_2,\sigma^2_2)Q(\delta_2,\sigma^2_2)d\delta_2.
\label{s2b}
\end{equation}
If we now set the smaller scale to be that of the observed collapsed
halo, and the overdensity at that scale to be the critical overdensity
for spherical collapse, Equation (\ref{s2b}) will give the resulting
overdensity at any larger scale due to the presence of this massive
galaxy.  In other words, we set $\delta_1=\delta_c(z_g)$ and
$\sigma^2_1=\sigma^2(M)$ in Equation (\ref{s2b}).

The larger scale corresponds to the field of observation.  In order to
calculate that, we first note that the excursion set principle
functions entirely in Lagrangian coordinates.  As a region evolves
towards eventual collapse its Lagrangian size stays unchanged while
its Eulerian size changes.  For a spherical region the Eulerian
evolution will follow the solution of the spherical collapse model.
However, since the Eulerian and Lagrangian sizes of the region
coincide at the initial instant, the spherical collapse solution is
also a relationship between these two sizes.  Thus we have
\begin{equation}
  R_E=\frac{3}{10}\frac{1-\cos\theta}{\delta_L}\frac{D(z=0)}{D(z)}R_L,
\label{sphcol}
\end{equation}
where $\theta$ is a parameter given by 
\begin{equation}
  \frac{1}{1+z}=\frac{3\times
    6^{2/3}}{20}\frac{(\theta-\sin\theta)^{2/3}}{\delta_L}.
\end{equation}
In our case, the Eulerian size of the region of interest is just the
comoving distance corresponding to the angular field of view, which is
just the angular diameter distance at the relevant redshift multiplied
by the angular field of view. 
The WFC3/IR field is $136^{''}\times123^{''}$.  For the best fit
$\Lambda$CDM cosmology the diagonal size of this field corresponds to
a comoving Eulerian distance $R_E=1.365$ Mpc at $z=8$.
In a WFC3/IR field centred on the object UDFy-42886345 at redshift
$8.0$ and apparent magnitude $H_{160,AB}=28.0$ we obtain a halo mass
$M=2.52\times 10^{11}$ M$_\odot$ and luminosity $L_{1500}=2.21\times
10^{29}$ erg s$^{-1}$ Hz$^{-1}$.

Notice, however, that since $\delta_L$ is unknown, Equation
(\ref{sphcol}) implies that the relation between the Eulerian size
$R_E$ and Lagrangian size $R_L$ is not one-to-one.  Thus, for the
probability distribution of linearly extrapolated overdensity $\delta$
given the halo mass $M$, we can only write 
\begin{equation}
  \frac{dP(\delta|M)}{d\delta}\propto
  Q[\delta,R_L(\delta,R_E)|\delta_c(z),R(M)],
  \label{PDProbDis}
\end{equation}
where the constant of proportionality is calculated by using the
normalization condition $\int[dP(\delta|M)/d\delta]d\delta=1$.  

In our calculations, we work with the value of $\delta$ for which
$dP(\delta|M)/d\delta$ is maximum.  For the WFC3/IR field at $z=8$,
this turns out to be $\delta=8.86$ (linearly extrapolated to $z=0$),
which results in a Lagrangian size $R_L=1.482$ Mpc for the region of
interest.  Notice that since $\delta>\delta_c$ the region must have
collapsed at some redshift $z\lesssim 6.5$.

In order to incorporate this overdensity into our reionization model,
note first that the number density of collapsed objects in such
overdense regions is enhanced with respect to that in a region with
average density.  This enhancement can be calculated using the
excursion set formalism \citep{1991ApJ...379..440B}.  It is then shown
in the Appendix that the comoving number density $N(M,z,z_c)dMdz_c$ at
redshift $z$ of collapsed halos having mass in the range $M$ and
$M+dM$ and redshift of collapse in the range $z_c$ and $z_c+dz_c$ is
given in this case by
\begin{equation}
\begin{split}
N(M,z,z_c)dMdz_c =
N(&M,z_c)\left(\frac{\nu^2\delta_c}{\delta_c/D(z_c)-\delta}\right)\frac{\dot 
  D(z_c)}{D^2(z_c)}\\&\times
p_\mathrm{surv}(z,z_c)\frac{dt}{dz_c}dz_cdM,
\end{split}
\label{nmzzc_biased}
\end{equation}
where $N(M,z_c)$ is the PS mass function.  The scale $R$ enters via the
definition of $\nu(M,z_c)$, which is now given by
\begin{equation}
\nu(M,z_c)=\frac{\delta_c/D(z_c)-\delta}{\sqrt{\sigma^2(M)-\sigma^2_R}}.
\end{equation}
The survival probability $p_\mathrm{surv}(z,z_c)$ is given by
\begin{equation}
p_\mathrm{surv}(z,z_c)=\frac{\delta_c/D(z)-\delta}{\delta_c/D(z_c)-\delta}.
\end{equation}

Another change when our reionization model is applied to overdense
regions is that we now normalize the probability distribution of
inhomogeneities in the IGM such that the average density in the region
is $\Delta=\delta+1$.

\begin{figure*}
\begin{center}
\begin{tabular}{cc}
\includegraphics[scale=0.7]{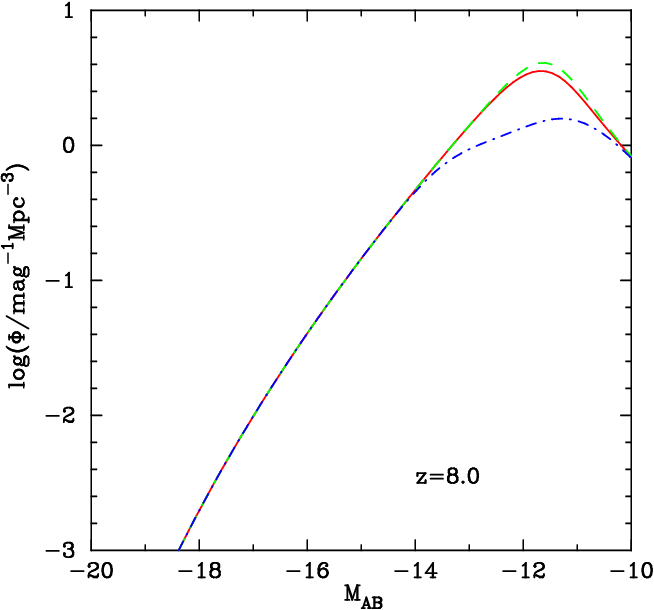} 
\includegraphics[scale=0.7]{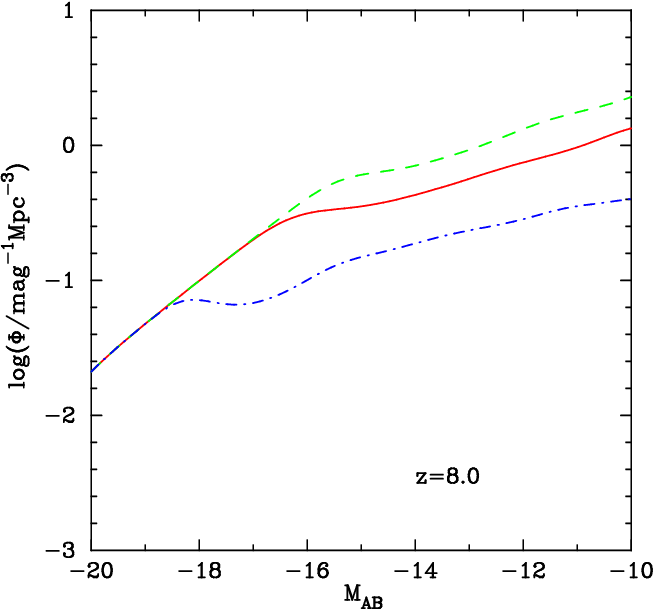} 
\end{tabular}
\end{center}
\caption{Effect of reionization history on luminosity function at
  z=8.0.  Solid, dashed and dot-dashed lines have $\tau=$ 0.073, 0.058
  and 0.088 respectively.  Left panel shows the average case.  Right
  panel shows the overdense case.}
\label{lf_reion}
\end{figure*}

\subsection{Radiative feedback}
\label{radfb}

As we argue in the next section, the luminosity function of galaxies
in an overdense region could carry an enhanced signature of feedback.
We therefore highlight our feedback model in this subsection.

Radiation from stars in the first galaxies is expected to ionize and
heat the surrounding medium.  This increases the mass scale above
which baryons can collapse in haloes.  Also, as a result, the minimum
mass of haloes that are able to cool is much higher in ionized regions
than in the neutral ones.  In our calculations, feedback appears
through the quantity $M_\mathrm{min}(z)$ in Equation \ref{global_sfr}.
The temperature evolution of both regions is calculated
self-consistently.  In the ionized regions, we fix the cut-off mass to
that corresponding to a virial temperature of $10^4$ K or the local
Jeans mass, whichever is higher.  In the neutral regions, since the
Jeans mass is always low, the cut-off mass always corresponds to the
virial temperature of $10^4$ K.  The minimum mass corresponds to the
circular velocity of 
\begin{equation}
  v_c^2=\frac{2k_\mathrm{boltz}T}{\mu m_p},
\end{equation}
where $\mu$ is the mean molecular weight.  For a temperature of
$\approx 10^4$ K, the minimum circular velocity is $\approx 25$ km
s$^{-1}$.  Note that this value is comparable to values obtained in
simulations \citep{2000ApJ...542..535G} but is somewhat higher than
that taken in the semi-analytic prescription of
\citet{2007MNRAS.377..285S}.

\begin{figure}
  \begin{center}
  \includegraphics[scale=0.65]{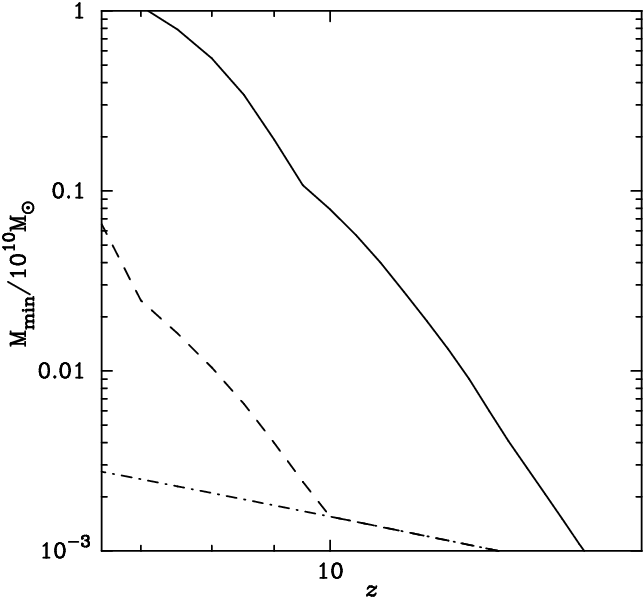} 
  \end{center}
  \caption{Evolution of the minimum mass $M_\mathrm{min}(z)$ of haloes
    that can host galaxies in ionized regions in the average case
    (dashed line) and the overdense case (solid line).  The dot-dashed
    line shows the minimum mass in neutral regions, which is same for
    the average and overdense cases.}
  \label{mmin} 
\end{figure}

We find that $M_\mathrm{min}(z)$ increases with time taking values of
$\approx 10^7$ M$_\odot$ at $z\approx 10$ and $\approx 10^8$ M$_\odot$
at $z\approx 7$.  In overdense regions the minimum mass is enhanced to
about $10^{10}$ M$_\odot$.  Figure \ref{mmin} shows the evolution of
the minimum mass.
\section{Results}

The results for reionization and thermal histories within overdense regions
are presented in this section.

\subsection{Effect of overdensity on reionization history}

We first consider the effect of overdensity on reionization history
for our fiducial model.  As is well known, reionization proceeds
differently in overdense regions.  The solid lines in Figure
\ref{BestFitReionizationModel} show the evolution of the
photoionisation rate, temperature in ionised regions, star formation
rate density and the volume filling factor of ionised regions in an
overdense region with size $R_L=1.482$ Mpc and linearly extrapolated
overdensity $\delta=8.86$.  This corresponds to the HUDF WFC3/IR field
centred at the brightest source in \citet{2010ApJ...709L.133B}.  (See
Section \ref{biassec}.)  Clearly while the average region is
completely ionised at $z\approx 6$, the biased region is ionised much
earlier, at $z\approx 7.5$.  This result agrees with
\citet{2007MNRAS.375.1034W}, although note that unlike that work, here
we calculate the clumping factor from a physical model for
inhomogeneities.
The reason for early reionization in overdense regions is the
enhanced number of sources, which is clear from the plots of 
photoionisation rate and the star-formation rate, both
of which are $\sim 5$ times higher than the corresponding 
globally averaged values. However, these overdense regions have
more recombinations, which results in enhanced temperatures as
is clear from the top right panel.
This results in enhanced negative radiative feedback which 
will suppress star formation in low mass galaxies and hence affect
the shape of the luminosity function.
In fact, for the average case, haloes in ionised regions
with masses below $10^8$ M$_\odot$ cannot form stars, whereas this
cutoff mass rises to close to $10^{10}$ M$_\odot$ in the overdense
case. Clearly feedback is enhanced in overdense regions.

\subsection{Effect of overdensity on luminosity function}

We now discuss the effect of overdensity on luminosity function.
Clearly, overdense regions have enhanced number of sources, hence
it is natural that the amplitude of the
luminosity function for such regions should be
higher than the globally averaged values.
However, the overdense regions have enhanced radiative feedback too, and hence
we expect a decrease in the number of sources, particularly towards
the fainter end.

Figure \ref{lf78} shows the effect of
overdensity on the luminosity function at $z=8$ for our fiducial
model.  The ionised volume filling factor within the overdense region
is $Q \approx 1.0$ for the overdense region under consideration at
this redshift.  The average region luminosity function (dashed line)
is clearly very different from the luminosity function in the
overdense region (solid line) at that redshift.  Firstly, we can
clearly see an enhancement in the source counts for brighter galaxies,
which is as expected. In addition, there is a clear sign of a
flattening  for magnitudes $M_{\rm AB} \gtrsim -17$, which is a signature
of radiative feedback. In comparison, the effect of feedback for
average regions occurs at much fainter magnitudes $M_{\rm AB} \sim
-12$. Note that there is no complete suppression of star formation 
for halo masses lower than the feedback threshold, rather
the luminosity function for magnitudes below the knee continues to
grow in the flattened region.  This is simply due to the continued
star formation in haloes with mass less than the cutoff mass at
$z=8.0$, but which collapsed at higher redshifts when the feedback
threshold mass was lower.  Thus, for instance,
if star formation is allowed to happen in a halo for only for a
fraction of the dynamical time [see equation (\ref{halo-sfr})], 
the luminosity function will rise less
steeply at the fainter end.  For small enough star formation time
scale, the luminosity function will show an abrupt cutoff.  Of course,
an abrupt cutoff is always seen at low enough luminosities, which are
not shown in the figure here.

It is important to understand here that the data points in Figure
\ref{lf} do not represent luminosity function of the overdense region.
Instead, those data points represent the globally averaged luminosity
function derived using a maximum likelihood procedure from the
observed luminosity distribution of sources.  In this procedure, a
likelihood function is defined, which describes the step-wise shape of
the luminosity function that is most likely given the observed
luminosity distribution in the search fields.  Details of this
procedure are described, for example, in Section 5.1 of
\citet{2010arXiv1006.4360B} and references therein.

\subsection{Luminosity function as a probe of reionization}

Given the fact that the effect of radiative feedback shows up at
brighter magnitudes for overdense regions, it is possible to use this
feature for studying feedback using near-future observations.  For
this purpose, we consider two additional models (other than the
fiducial one) of reionization.  These models have parameter values
($f_*$, $f_{\rm esc}$) = ($0.06$, $0.3$) and ($0.2, 0.07$) and we
obtain $\tau_e = 0.088$ and $0.058$ respectively for these models.  We
fix $f_*$ and only change the value of $f_{\rm esc}$ to ensure that
any effect on the luminosity function is purely due to feedback.
These two models predict photoionisation rates greater and lesser
respectively than what are presented by \citet{2007MNRAS.382..325B}.

The right panel of Figure \ref{lf_reion} shows the luminosity function
at $z=8$ within the overdense region for three different reionization
histories, which can be compared with the corresponding luminosity
function in average region (shown in the left panel).  In both cases a
distinct ``knee'' is seen in the luminosity function as a signature of
feedback.  The luminosity function flattens at this luminosity, and is
suppressed to very low values at much lower luminosities.  As
described in the previous section, this signature of feedback
appears at brighter magnitudes for the overdense region.  This is
expected, because the cutoff mass depends directly on the temperature,
which is enhanced in the overdense region.  We also note that in the
case of the first model the flattening occurs for $M_{\rm AB} \gtrsim
-19$ whereas for the second model at a fainter luminosity of $M_{\rm
  AB} \simeq -16$.  This is due to the fact that the photoionisation
feedback is enhanced in the first model due to enhanced flux.

The evolution of the filling factor affects this result through the
average temperature which sets the cutoff mass.  Thus, early and late
reionization models are distinguished by the difference in the nature
of flattening in both cases.  This also affects the evolution of the
luminosity function.

We find that the reionization history has a strong effect on the
luminosity function at the faint end.  It is known that the bright end
of the luminosity function is affected primarily by the star formation
mode of a halo, and the overall bias, whereas its faint end is
affected by the reionization history.

However, we also find, from Figure \ref{lf_reion}, that the effect of
reionization history is much stronger in the case of overdense
regions.  This is because of the enhanced photoionisation feedback,
which is more sensitive to changes in reionization history.  This
order of magnitude change in the overdense region luminosity function
should be visible to the James Webb Space Telescope, which can observe
up to $m_{\rm AB}\approx 31.5$ ($M_{\rm AB} \approx -16.0$ at
redshifts of interest; \citealt{2006AAS...20921007W}).

\section{Discussion and Summary}

We have used a semi-analytic model, based on
\citet{2005MNRAS.361..577C, 2006MNRAS.371L..55C} to study reionization
and thermal history of an overdense region. Studying such regions is
important because observations of galaxy luminosity function at high
redshifts typically focus fields of view of limited sizes
preferentially containing bright sources; these regions possibly are
overdense and hence biased with respect to the globally averaged
regions. In particular, we study the effect of radiative feedback
arising from reionization on the shape of galaxy luminosity function.

In summary, we find that 
\begin{enumerate}
\item Reionization proceeds differently in overdense regions.
  Overdense regions are ionised earlier because of enhanced number of
  sources and star formation. In addition, these regions have higher
  temperatures because of enhanced recombinations and hence effect of
  radiative feedback is enhanced too.

\item In particular, the shape of the galaxy luminosity function for
  biased regions is very different from that for average
  regions. There is a significant enhancement in the number of
  high-mass galaxies because of bias, while there is a reduction in
  low-mass galaxies resulting from enhanced radiative feedback.

\item Luminosity function in overdense regions is more sensitive to
  reionization history compared to average regions.  The effect of
  radiative feedback shows up at absolute AB magnitudes $M_{\rm AB}
  \gtrsim -17$ in these regions, while it occurs at much fainter
  magnitudes $M_{\rm AB} \sim -12$ for average regions.  This order of
  magnitude change in the overdense region luminosity function should
  be visible to the James Webb Space Telescope in future.
\end{enumerate}

Finally, we critically examine some of the simplifying assumptions
made in this work and how they are likely to affect our conclusions.
Firstly, we have seen that the presence of a high mass galaxy within a
region if size $R$ does not uniquely specify the value of the
overdensity $\delta$. Rather we obtain a probability density (which is
Gaussian in shape) and work with the value where this probability is
maximum.  In reality, however, the actual value of $\delta$ could be
different and this may possibly affect the predicted luminosity
function. Note that the luminosity function at the brighter end is
almost independent of the details of reionization history, and this,
in principle, can be used for constraining the value of $\delta$. The
effect of feedback can then be studied using the faint end of the
luminosity function.

The radiative feedback prescription used in this paper is based on a
Jeans mass calculation \citep{2005MNRAS.361..577C}. However, alternate
prescriptions for feedback exist in literature, e.g.,
\citet{2000ApJ...542..535G} and hence the shape of the luminosity
function at faint ends as predicted by our model may not be robust.
Interestingly, the presence of a ``knee'' in the luminosity function
can be used to estimate the value of the halo mass below which star
formation can be suppressed (which in turn can indicate the
temperature) while the shape of the function below this knee should
indicate the nature of feedback. This study can also be complemented
with proposed for studying feedback using other observations, e.g., 21
cm observation \citep{2008MNRAS.384.1525S} and CMBR
\citep{2008MNRAS.385..404B}.

Finally, we have neglected the presence of other sources of
reionization, e.g., metal-free stars, minihaloes, and so on.  It is
expected that these sources would be too faint to affect the
luminosity function in the ranges we are considering. However, these
sources may affect the thermal history of the medium, e.g, the
metal-free stars would produce higher temperatures because of harder
spectra. In such cases, it is most likely that feedback would occur at
magnitude brighter than what we have indicated and hence would
possibly be easier to detect.

\section*{Acknowledgements}

GK acknowledges useful discussion with Prof.~Jasjeet S. Bagla.
Computational work for this study was carried out at the cluster
computing facility in the Harish-Chandra Research Institute
(http://cluster.hri.res.in/index.html).  We would also like to thank
the referee for suggestions that improved this paper's quality.

\bibliographystyle{mn2e}
\bibliography{fbck}

\section*{Appendix: Formation rate and survival probability of haloes in overdense regions} 

As expressed in Equation (\ref{nnu}), the number density of ionizing
photons produced per unit time is related to the SFR density, which in
turn depends on the SFR in each halo, given by Equation
(\ref{global_sfr}), and the number density of haloes of a certain age,
given by Equation (\ref{nmzzc}) for average regions, and by Equation
(\ref{nmzzc_biased}) for overdense regions.  We derive Equation
(\ref{nmzzc_biased}) in this appendix.

We denote the number density at redshift $z$ of haloes formed between
redshifts $z_c$ and $z_c+dz_c$, with mass between $M$ and $M+dM$, by
$N(M,z,z_c)dMdz_c$.  This quantity is related to (1) the formation
rate at redshift $z_c$ of haloes with mass between $M$ and $M+dM$,
denoted by $\dot N_\mathrm{form}(M,z_c)dM$, and (2) the probability of
their survival at redshift $z$, denoted by $p_\mathrm{surv}(z,z_c)$.
We calculate these two quantities using a technique given by
\citet{1994PASJ...46..427S}, applied to an overdense region with
overdensity $\delta$ and size $R$.

Recall that in extended Press-Schechter theory
\citep{1991ApJ...379..440B}, the mass function of dark matter haloes
is defined as the comoving number density of haloes with mass between
$M$ and $M+dM$.  At redshift $z$, this quantity is given by
\begin{equation}
  N(M,z)dM=\sqrt{\frac{2}{\pi}}\frac{\bar\rho_m}{M}\exp\left(\frac{-\nu^2}{2}\right)\frac{d\nu}{dM}dM,
  \label{ps-mf}
\end{equation}
where $\bar\rho_m$ is the average matter density, and, as before,
$\nu(M,z)\equiv\delta_c/[D(z)\sigma(M)]$.  The critical overdensity of
collapse of a halo is denoted by $\delta_c$, $D(z)$ is the growth
function of density perturbations, and $\sigma(M)$ is the rms value of
density perturbations at the comoving scale corresponding to mass $M$.
In a region with overdensity $\delta$ and linear size $R$, the mass
function is enhanced.  This enhancement can be calculated using the
excursion set formalism \citep{1991ApJ...379..440B}.  The resulting
mass function is again given by Equation (\ref{ps-mf}), except that
now the quantity $\nu(M,z)$ is defined as
\begin{equation}
  \nu(M,z)\equiv\frac{\delta_c/D(z)-\delta}{\sqrt{\sigma^2(M)-\sigma^2_R}},
\end{equation}
where $\sigma_R$ is the rms value of density perturbations at comoving
scale $R$.  Closely following \citet{1994PASJ...46..427S}, we can
write
\begin{equation}
  \dot N(M,z)=\dot N_\mathrm{form}(M,z)-\dot N_\mathrm{dest}(M,z),
\end{equation}
where $\dot N_\mathrm{dest}(M,z)dM$ is the destruction rate at
redshift $z$ of haloes of mass between $M$ and $dM$.  (The halo
formation rate is defined as the number density of haloes formed per
unit time from mergers of lower mass haloes.  Similarly the halo
destruction rate is defined as the number density of haloes destroyed
per unit time due to mergers with other haloes.)  Here, an overdot
denotes the time derivative.  We can write the destruction rate as
\begin{eqnarray}
\dot N_\mathrm{dest}(M,z)&=&\int_M^\infty N(M,z)\tilde Q(M,M^\prime,z)dM^\prime,\\
&\equiv&\phi(M,z)N(M,z),
\label{probqtilda}
\end{eqnarray}
and the formation rate as 
\begin{equation}
\dot N_\mathrm{form}(M,z)=\int_{M_\mathrm{min}}^M N(M^\prime,z)Q(M^\prime,M,z)dM^\prime,\\
\label{probq}
\end{equation}
where $\tilde Q(M,M^\prime,z)$ is the probability that a halo of mass
$M$ merges with another halo to result in a halo of mass $M^\prime$
per unit time, and $Q(M^\prime,M,z)$ that an halo of mass $M$ forming
at redshift $z$ has a progenitor of mass $M^\prime$.  The threshold
mass $M_\mathrm{min}$ is introduced at this stage to avoid divergence.
This gives
\begin{equation}
\dot N_\mathrm{form}(M,z)=\dot N(M,z)+\phi(M,z)N(M,z).
\label{ndotform}
\end{equation}
We now assume that $\phi$ has no characteristic mass scale so that
$\phi(M,z)=M^\alpha\tilde\phi(z)$.  This gives 
\begin{equation}
  \tilde\phi(z)=\frac{-\dot N(M,z)+\dot N_\mathrm{form}(M,z)}{N(M,z)M^\alpha}.
  \label{phi1}
\end{equation}
But since the left hand side of Equation (\ref{phi1}) is a function of
time alone (through the redshift), the right hand side of this
equation also has to be independent of mass.  In particular, we can
then set $M=M_\mathrm{min}$ in this equation, giving us
\begin{equation}
  \tilde\phi(z)=\frac{-\dot N(M_\mathrm{min},z)}{N(M_\mathrm{min},z)M_\mathrm{min}^\alpha}.
\end{equation}
Now, in the case of the overdense region that we are considering here,
we have
\begin{equation}
  \dot N(M,z)=N(M,z)\frac{\dot D(z)}{D^2(z)}\frac{\delta_c}{\delta_c/D(z)-\delta}[\nu^2(m,z)-1],
\end{equation}
which gives 
\begin{equation}
  \tilde\phi=\frac{\dot D}{D^2}\frac{\delta_c}{\delta_c/D(z)-\delta}[\nu^2(M_\mathrm{min},z)-1]M_\mathrm{min}^{-\alpha}.
\end{equation}
Since our choice of threshold mass $M_\mathrm{min}$ is arbitrary, we
now need to take the limit $M_\mathrm{min}\rightarrow 0$.  However,
since $\nu\rightarrow 0$ in this limit, $\tilde\phi$ becomes
indeterminate, except when $\alpha=0$.  This implies that we must set
$\alpha=0$ for consistency.  This gives $\phi(M,z)=\tilde\phi(z)$.
Substituting the resultant expression in Equation (\ref{ndotform}), we
get
\begin{equation}
  \dot N_\mathrm{form}(M,z)=N(M,z)\frac{\dot D}{D^2}\frac{\delta_c}{\delta_c/D(z)-\delta}\nu^2(M,z).
  \label{nformdot_ovd}
\end{equation}
This the required formation rate of haloes in an overdense region.

Furthermore, from our definitions of probabilities in Equations
(\ref{probqtilda}) and (\ref{probq}), we can write the probability
that a halo that has formed at redshift $z_c$ continues to exist at
redshift $z$ as
\begin{equation}
  p_\mathrm{surv}(z,z_c)=\exp\left[-\int_{t(z_c)}^{t(z)}\phi(t^\prime)dt^\prime\right],
\end{equation}
which in our case results in 
\begin{equation}
  p_\mathrm{surv}(z,z_c)=\frac{\delta_c/D(z)-\delta}{\delta_c/D(z_c)-\delta}. 
  \label{psurv_ovd}
\end{equation}
From Equations (\ref{nformdot_ovd}) and (\ref{psurv_ovd}), we can now
write the the comoving number density $N(M,z,z_c)dMdz_c$ at redshift
$z$ of collapsed halos having mass in the range $M$ and $M+dM$ and
redshift of collapse in the range $z_c$ and $z_c+dz_c$ as
\begin{equation}
\begin{split}
N(M,z,z_c)dMdz_c =
N(&M,z_c)\left(\frac{\nu^2\delta_c}{\delta_c/D(z_c)-\delta}\right)\frac{\dot 
  D(z_c)}{D^2(z_c)}\\&\times
p_\mathrm{surv}(z,z_c)\frac{dt}{dz_c}dz_cdM,
\end{split}
\end{equation}
This is our Equation (\ref{nmzzc_biased}).  

It is worth pointing out that Equations (\ref{nformdot_ovd}) and
(\ref{psurv_ovd}) reduce to the average forms for halo formation rate
and survival probability in the limit $\delta\rightarrow 0$ and
$R\rightarrow\infty$.
\end{document}